\documentclass[]{aa} 

\usepackage{graphicx} 
\usepackage{amssymb} 
\usepackage{epsfig} 
\usepackage{natbib} 
 
\def\pl{{\sc pl}}

\def\bb{{\sc bb}}

\def\bb{{\sc bb}} 
\def\I{{\em INTEGRAL}} 
\def\IGR{IGR~J17254-3257} 
\def\be{\begin{equation}} 
\def\ee{\end{equation}}

\begin{document} 
 
\title{IGR~J17254-3257, a new bursting neutron star} 
 
\author{J. Chenevez\inst{1} 
   \and M. Falanga \inst{2,3}  
   \and E. Kuulkers \inst{4}  
   \and R. Walter \inst{5,6} 
   \and L. Bildsten \inst{7}
   \and S. Brandt \inst{1} 
   \and N. Lund \inst{1} 
   \and T.~Oosterbroek \inst{8} 
   \and J. Zurita Heras\inst{2}
}

\offprints{J. Chenevez} 
\titlerunning{IGR~J17254-3257, a new bursting neutron star} 
\authorrunning{J. Chenevez, et al.} 
 
\institute{Danish National Space Center, Technical University of Denmark, 
           Juliane Maries Vej 30, 2100 
           CopenhagenØ, Denmark \email{jerome@spacecenter.dk}
\and CEA Saclay, DSM/DAPNIA/Service d'Astrophysique 
     (CNRS FRE 2591), 91191, Gif sur Yvette, France 
\and Unit\'e mixte de recherche Astroparticule et 
     Cosmologie, 11 place Berthelot, 75005 Paris, France 
\and ISOC, ESA/ESAC, Urb. Villafranca del Castillo, 
     P.O. Box 50727, E-28080 Madrid, Spain 
\and INTEGRAL Science Data Centre, Chemin d'\'Ecogia 16, 1290 Versoix, Switzerland
\and Observatoire de Gen\`eve, Universit\'e de Gen\`eve, 51 ch. des Maillettes, 
     1290 Sauverny, Switzerland 
\and Kavli Institute for Theoretical Physics, Kohn Hall, University of 
     California, Santa Barbara, CA 93106
\and Science Payload and Advance Concepts Office, ESA-ESTEC, Postbus 299, 
     2200 AG, Noordwijk, The Netherlands  
}

\abstract 
{} 
{
The study of the observational properties of uncommonly long bursts 
from low luminosity sources 
is important when investigating the transition from a hydrogen\,-\,rich bursting 
regime to a pure helium regime and from helium burning to 
carbon burning as predicted by current 
burst theories. 
On a few occasions 
X-ray bursts have been  
observed with extended decay times up to several tens of minutes,  
intermediate between usual type I X-ray bursts 
and so-called superbursts. 
}  
{
\IGR\ is a recently discovered X-ray burster of which only two bursts have  
been recorded: 
an ordinary short type I X-ray burst, 
and a 15~min long burst. 
The properties of the X-ray bursts observed
from \IGR\ are investigated.  The broad-band spectrum of the
persistent emission in the 0.3--100 keV energy band is studied using
contemporaneous {{\it INTEGRAL}} and {{\it XMM-Newton}} data.  
}   
{
A refined position of \IGR\ is given and an upper limit to its distance 
is estimated to about 14.5 kpc. 
The persistent bolometric flux of 
$1.1\times10^{-10}$ erg cm$^{-2}$ s$^{-1}$ 
corresponds, at the canonical distance of 8~kpc, to
$L_{\rm pers}\approx~8.4\times 10^{35}$ erg s$^{-1}$ 
between 0.1--100 keV, 
which translates to a mean accretion rate of about 
$7\times 10^{-11} M_{\odot} yr^{-1}$.
} 
{
The low X-ray persistent luminosity of \IGR\ seems to indicate the 
source may be in a state of low accretion rate usually associated with 
a hard spectrum in the X-ray range. 
The nuclear burning regime may be intermediate between pure He and 
mixed H/He burning.
The long burst is the result of the accumulation of a thick He layer,
while the short one is a prematurate H-triggered He burning burst
at a slightly lower accretion rate.
}

\keywords{binaries: close -- stars: individual: 
IGR~J17254-3257 = 1RXS~J172525.5-325717 -- stars: neutron 
-- X--rays: bursts} 
 
\maketitle 
 
\section{Introduction} 
\label{sec:intro} 
 
The X-ray source \IGR\ was discovered with the International Gamma-Ray 
Astrophysics Laboratory (\I) in the Galactic Centre hard X-ray survey 
and was reported at $11\sigma$ detection 
in a 20--60 keV mosaic accumulated between February 27 and October 19, 2003 
\citep{walter04}.  
From the derived source coordinates, \IGR\ was tentatively identified as 
the {\it ROSAT} source 1RXS~J172525.5-325717 
\citep{stephen05}. However, little was known about the nature of the 
compact source and the system. An ordinary type I X-ray burst 
which occurred on February 17, 2004 with a peak flux of 0.8 Crab 
at 3--30 keV has been detected in \I/JEM-X archival data 
\citep{brandt06}.  
 
We found a second burst of \IGR\ lasting about 
fifteen minutes 
in \I\ observations of the Galactic Centre 
performed on October 1st 2006. 
Only on a few occasions have type I X-ray bursts shown decay times ranging 
between ten and a few tens of minutes.
Known examples of such observations are 
4U~1724-307 \citep{Swank},
4U~1708-23 \citep[e.g.,][]{Hoff},
GX~17+2 \citep{kul02},
SLX~1737-282 \citep{intZ02},
SLX~1735-269 \citep{molkov05}, 
2S~0918-549 \citep{intZ05}, and 
GX~3+1 \citep{chenevez06}.  
These long bursts have durations and energy releases ($\sim\,10^{41}$ erg) 
intermediate between usual type I X-ray bursts and so-called superbursts 
which last more than an hour \citep[e.g.,][]{ek04}. 
The mechanisms driving such long bursters at very low persistent luminosity 
have been the subject of 
recent investigations 
\citep[e.g.,][]{Peng, CoopN}, suggesting that thermally unstable hydrogen 
ignition results in sporadic energetic helium bursts in a mixed hydrogen 
and helium environment. However, long helium bursts are also observed 
at low pure helium accretion rates \citep[e.g.,][]{intZ05, intZ07}. 
\IGR\ thus provides a new example of these few faint persistent neutron stars 
(NS) that display type I X-ray bursts of such different durations.    
 
In this {\em letter} we report the \I/JEM-X results of the light curve 
and spectral analysis of the short and long type I bursts of \IGR, 
hereafter referred to as burst 1 and  burst 2, respectively. 
Using {\em XMM-Newton} (0.3--10 keV) and \I\ (5--100 keV) data we also study 
the broad-band spectrum of the persistent emission of this source.

\section{Data Analysis and results} 
\label{sec:integral} 
 
Burst 1 and burst 2
have been discovered in \I/JEM-X 
\citep{w03,lund03} 3--20 keV data, on February  17, 2004 
and  October 1st,  2006, respectively. 
In the IBIS/ISGRI \citep{u03,lebr03} 18--40 keV energy band 
neither of the two bursts were detected at a statistically significant level.

The persistent broad-band emission (0.3--100 keV) of \IGR\ is obtained from 
\I/JEM-X (3--20 keV) and IBIS/ISGRI (20--100 keV) data from  
March 2003 to October 2006, 
as well as {\em XMM-Newton} \citep{j01} data acquired on April 2, 2006 
(MJD 53827) for about 10 ks.
The \I\ data were extracted for all pointings within $3^{\circ}$
(JEM-X) and $4^{\circ}$ (ISGRI) of the source position 
for a total effective exposure (taking account of instrumental effects) 
of about 400~ks and 830~ks, respectively. 
To study the (weak) persistent X-ray emission, the JEM-X and ISGRI
spectra have been derived from mosaic images in six energy bands 
for JEM-X (3--20 keV) and four energy bands for ISGRI (20--100 keV). 
The data reduction was performed using the standard 
Offline Science Analysis (OSA) software version 6.0.  
A systematic error of 2~\% was applied to JEM-X and ISGRI spectra, 
which corresponds to the current uncertainties in the response matrices.
All uncertainties in the spectral parameters are given at a 90~\% 
confidence level for single parameters.

We used 0.3--12 keV  {\it XMM-Newton} data from the EPIC -pn \citep{str01} 
and -MOS1 cameras \citep{t01} operating in full window mode. 
The data reduction and analysis were done with the 
{\em XMM-Newton} Science Analysis System (SAS version 7.0).
These data allowed us to derive a refined source position at
$\alpha_{\rm J2000} = 17^{\rm h}25^{\rm m}24\fs8$ and 
$\delta_{\rm J2000} = -32{\degr}57\arcmin15\arcsec$ with an
uncertainty of $2\arcsec$. This is still inside the {\it ROSAT} error box 
of 1RXS~J172525.5-325717, 
but it excludes the 
optical counterpart suggested by \citet{stephen05}; 
no other known optical or infrared counterpart is found in 
the {\em XMM-Newton} error circle.

\subsection{Persistent emission}  
 \label{sec:persitent} 

\begin{figure}[htb] 
\centerline{\epsfig{file=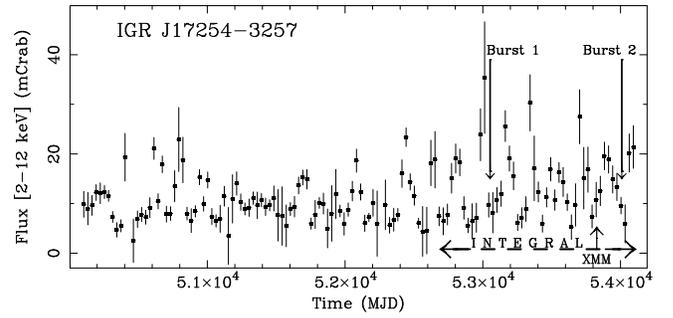,angle=-90,width=8.5cm}} 
\caption 
{{\it RXTE}/ASM light curve for \IGR\ averaged over 30-day  
intervals from January 6, 1996.   
The count rate has been converted into flux  
using 1 Crab Unit for 75 cts s$^{-1}$ \citep{l96}.
The times of burst 1 and burst 2, as well as the present data set coverage, 
are indicated by arrows.} 
\label{fig:asm} 
\end{figure}

In Fig. \ref{fig:asm} we show the 2--12 keV persistent emission for 
\IGR\ obtained with 
{\em RXTE}/ASM \citep{l96}. The source is a weak persistent X-ray source.
Since the persistent emission of \IGR\ is more or less stable over the whole 
available data set, its broad-band spectrum has been obtained by combining the
{\em XMM-Newton} and \I\ observations. 

\begin{figure} 
\centerline{\epsfig{file=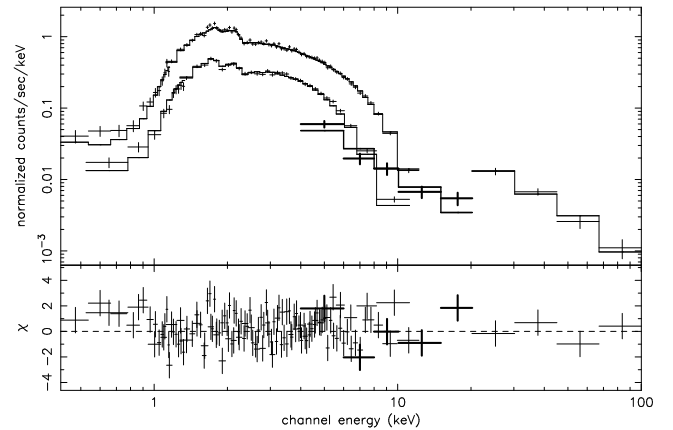,angle=-90,width=8.5cm}}
\caption{The persistent spectrum of \IGR\ fit with a photoelectrically-absorbed 
blackbody (\bb) plus a cutoff power-law (\pl) model. 
The data points correspond to the EPIC- pn (0.3--12 keV), 
MOS1 (0.5--10 keV), JEM-X (5--20 keV - bold), and 
ISGRI (20--100 keV) spectra, respectively. 
The lower panel shows the residuals between the data and the model.
}  
\label{fig:spec} 
\end{figure}

The joined pn/MOS1/JEM-X/ISGRI spectrum 
is best fit with a photoelectrically-absorbed blackbody (\bb) and cutoff 
power-law (\pl) model (Fig. \ref{fig:spec}).
A multiplicative factor for each instrument was included in the fit to 
take account of the uncertainty in the cross-calibration of the instruments. 
The factor was fixed at 1 for the {\it XMM-Newton}/EPIC-pn data and the 
normalizations of the MOS1, JEM-X and ISGRI data were found within $0.89\pm0.04$. 
The best fit parameters, with $\chi^{2}{\rm /d.o.f.} = 716/697$, are: 
absorption $N_{\rm H}= 1.79^{+0.08}_{-0.1}\times10^{22}$ cm$^{-2}$, \bb\
temperature $kT_{\rm bb}=1.06^{+0.08}_{-0.1}$ keV, a \pl\ photon
index $\Gamma=1.64^{+0.07}_{-0.1}$, and cutoff energy
$E_{\rm c}=62^{+24}_{-24}$ keV. 
The 0.3--100 keV unabsorbed flux is $10^{-10}$ erg cm$^{-2}$ s$^{-1}$. 
Note that, if we replace the soft thermal emission by a  
multi-temperature disc \bb\ model \citep{mitsuda84}, we
find the same $\chi^{2}$ value with a higher $kT_{\rm bb}\approx~1.6$ keV, but
the inner disk radius, $R_{\rm in} \sqrt{\cos\,i} = 0.4^{+0.1}_{-0.1}$ km 
(at 8 kpc) is smaller than the expected NS radius.

\subsection{Burst light curves} 
\label{sec:lcburst} 
 
\begin{figure}[htb] 
\centerline{\epsfig{file=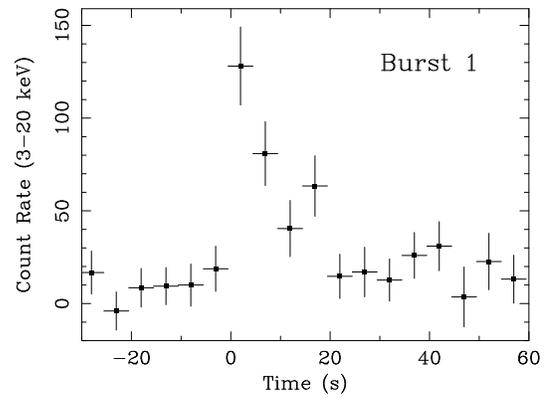,angle=-90,width=7.0cm}}
\caption 
{The short type I X-ray burst detected from \IGR\ on February 17, 2004.
Time 0 corresponds to 19:44:00 (UTC). 
The JEM-X (3--20 keV) net light curve is shown (background subtracted) 
with a time bin of 5 s.}  
\label{fig:burst1} 
\end{figure}

\begin{figure}[htb]
\centerline{\epsfig{file=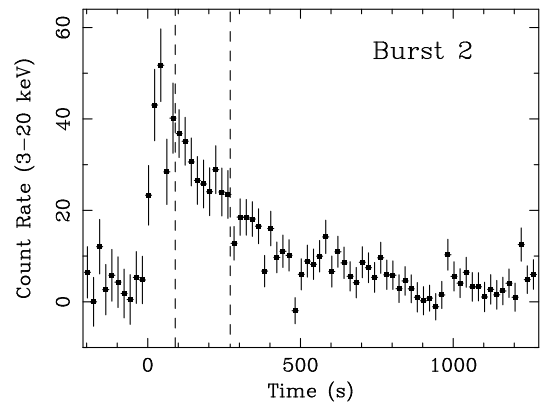,angle=-90,width=7.0cm}} 
\caption 
{Same as Fig. \ref{fig:burst1} for the long burst on October 1st, 2006 
with a time bin of 20 s. Time 0 corresponds to 7:13:37 (UTC).
The vertical dashed lines indicate the time of the \I\ slew interval.}  
\label{fig:burst2} 
\end{figure}

In Fig. \ref{fig:burst1} and \ref{fig:burst2} we show the  
JEM-X 3--20 keV light curves for bursts 1 and 2, respectively. 
Burst 2 was  observed during two consecutive stable pointings 
with a 2 minute satellite slew in between. 
Since the position of \IGR\ did not traverse the field of view, as well as 
the background due to other sources did not significantly evolve during 
the slew, it was possible to obtain the light curve during this interval 
by a simple linear interpolation of the detector event list.
Taking account of the slight illumination increase due to the apparent 
source position change between the two pointings, 
we renormalized the slew burst 
count rate to the background\,-\,subtracted light curve during the stable 
pointings for the same time bin and energy band. 
The slew time interval is shown in Fig. \ref{fig:burst2} between the 
vertical dashed lines.

Both bursts were soft and no emission was detected above 20 keV.
The start time for each burst was determined when the intensity rose to 10~\% 
of the peak above the persistent intensity level. 
The rise time is defined as the time between the start of the burst and 
the time at which the intensity reached 90~\% of the peak burst intensity.
For burst 1 it was $2\pm1$~s; the e-folding decay time was $15.7\pm6.7$~s 
and the total duration was about 29~s. 
Burst 2 was 
about half as intense as burst 1. 
However, its total duration  was at least 900~s 
with a rise time of $20\pm5$~s and an e-folding decay time of $219\pm32$~s. 
In both cases the persistent emission before and after the burst interval 
was not significantly detected in neither JEM-X nor ISGRI.

\subsection{X-ray burst spectra} 
\label{sec:burstspec} 
  
For the spectral analysis of the bursts we used JEM-X data in the 3--20~keV band.
For burst 1 we only performed a time-averaged (29 s) spectral analysis, 
due to its short duration. The net burst spectrum was well fit 
($\chi^{2}_{\rm red}=1.2$) by a simple \bb\ model. 
 
For the long burst 2 we analyzed the spectrum during the 95~s interval ('peak')
and the 645~s interval ('decay') before and after the slew, respectively.  
This allows to measure the thermal cooling during the decay, which is one 
of the charateristics of type I X-ray bursts.
Indeed, though the uncertainties of the spectral fits do not indicate 
a significant softening during the decay, 
the corresponding 3--6 keV / 6--20 keV hardness does decrease by a factor 2.
Finally, we determined the time-averaged (740~s) spectrum, excluding 
the slew interval. In each case, the same model as for burst 1 was used. 

The inferred \bb\ 
temperature, $kT_{\rm bb}$, and apparent \bb\ radius at 8~kpc, $R_{\rm bb}$, 
for both bursts are reported in Table \ref{table:spec1}.
The burst fluences are obtained from the bolometric fluxes, $F_{\rm bol}$, 
extrapolated in the 0.1--100 keV energy range and integrated over the 
respective burst durations.
The peak fluxes, $F_{\rm peak}$, are derived from the peak count rates 
in fine time resolution and renormalized for the same energy range.

\begin{table}[htb] 
\caption{Analysis results of the two bursts} 
\label{table:spec1}
\begin{center} 
\renewcommand{\footnoterule}{}
\begin{tabular}{lllll} 
\hline \hline
Dataset    & Burst 1 &      &  Burst 2  &   \\  \cline{3-5}
Parameters & average & peak & decay & average\\ 
\hline 
\noalign{\smallskip} 
$kT_{\rm bb}$ (keV) & 1.4$^{+0.5}_{-0.4}$  & 1.6$^{+0.3}_{-0.2}$ 
 &  1.2$^{+0.3}_{-0.2}$ & 1.3$^{+0.2}_{-0.2}$\\ 
$R_{\rm bb, d_{8 kpc}}$ (km) & 12$^{+13}_{-6}$ & 6.4$^{+3}_{-4}$ & 
5.6$^{+4}_{-2}$  &   5.1$^{+2}_{-2}$\\ 
$\chi^{2}/{\rm dof}$  &  12/10 & 48/49 &  48/42 & 59/47\\   
$F_{\rm bol}$ \small $^{a}$ & 8.9 & 4.9 & 1.0 & 1.1\\ 
\hline 
\multicolumn{2}{l}{Burst parameters} \\
$F_{\rm peak}$ \small $^{a}$   &   $\simeq~20$ &  & $\simeq~12$ \\
$f_{\rm b}$ \small $^{b}$   &   \multicolumn{1}{l}{$2.6 \times 10^{-7}$} &  & \multicolumn{2}{l}{$2.6 \times 10^{-6}$} \\
$\tau$ \small $^{c}$   &  13  &  & 216 \\
$\gamma$ \small $^{d}$  &  0.006  &  &  0.009 \\
\hline
\end{tabular}
\end{center} 
\small $^{a}$ Unabsorbed flux (0.1--100 keV) in units of $10^{-9}$erg cm$^{-2}$ s$^{-1}$. \linebreak
\small $^{b}$ Fluence (erg cm$^{-2}$).
\small $^{c}$ $\tau (sec) \equiv f_{\rm b}/F_{\rm peak}$.
\small $^{d}$ $\gamma \equiv F_{\rm pers}/F_{\rm peak}$; 
$F_{\rm pers}=1.1\times 10^{-10}$ erg cm$^{-2}$ s$^{-1}$ (0.1--100 keV).
\end{table}

\section{Discussion and conclusions}  
 \label{sec:discussion} 

We have found two X-ray bursts of very different durations from \IGR.
The first one is an ordinary short type I X-ray burst while the second one is 
similar to the other few intermediately long bursts reported to date.
Assuming that the bolometric peak luminosity of burst 1 was at most at 
the Eddington limit for a helium burst, 
$L_{\rm Edd}\approx\,3.8\times 10^{38}$ erg s$^{-1}$, 
as empirically derived by \citet{kul03},  
we can derive an upper limit to the distance of the source of about 14.5~kpc. 
This is also consistent with the high measured column density, 
which is a factor 2 higher than the total galactic column density 
as estimated from HI maps in that direction \citep{dickey90}.

The best fit to the broad-band 0.3--100 keV persistent emission spectrum of 
\IGR\ required a two-component model: 
a cutoff \pl\ together with a \bb\ soft component; the hard spectral
component contributes most to the observed flux (95~\%). 
These spectral characteristics are similar to those observed in the low/hard 
state of low-mass X-ray binaries (LMXB) weakly magnetized NS 
\citep[see, e.g.,][]{barret00, MF06}. 
The soft thermal emission, $kT_{\rm bb}$, could be
associated with the radiation from the hot-spot on the NS surface around 
an accretion shock. 
Assuming a canonical distance of 8 kpc for a LMXB in the direction of 
the Galactic Centre, the estimated persistent flux between 0.1--100 keV, 
$F_{\rm pers}=1.1\times 10^{-10}$ erg cm$^{-2}$ s$^{-1}$, 
translates to a bolometric luminosity 
$L_{\rm pers}\approx~8.4\times 10^{35}$ erg s$^{-1}$. 
This makes \IGR\ another member of the class of bursters with 
low persistent emission \citep[see, e.g.,][and references therein]{coc01, cor04}. 
The mass accretion rate per unit area of the persistent emission, 
given by $L_{\rm pers}\;\eta^{-1}$ c$^{-2}/
4\pi R_{\rm NS}^{2}$ 
(where $\eta \simeq~0.2$ is the accretion efficiency 
for a 1.4 M$_{\odot}$ and 10 km radius NS),
is $\dot m\approx~370$ g cm$^{-2}$ s$^{-1}$. 
Since the ASM light curve does not indicate strong differences in the persistent 
flux of the source at the time of the two bursts, we are not able to comment 
on the exact accretion state at any time. 
We assume therefore that both bursts occurred at 
about the same accretion rate. 
From the detection of only two bursts during 
the total observation time of about 860~ks elapsed on the source by JEM-X and 
{\em XMM-Newton}, we can estimate a recurrence time 
of five days.

Both bursts are 
well described by a simple \bb\ model 
representing the thermal emission from the neutron star surface, 
which is consistent with the observed properties of type I X-ray bursts 
\citep[see, e.g.,][]{galloway06}.
With a fast rise and exponential decay light curve 
burst 1 is indeed similar to commonly observed normal type I bursts. 
Its total energy release was $E_{b,1}\simeq 2\times 10^{39}$ erg (assuming 8 kpc 
distance), not that large considering the inferred long accumulation time. 
The 15 minute duration of burst 2 is more unusual. 
The total energy release $E_{b,2}\simeq 2\times 10^{40}$ erg (at 8 kpc distance)
corresponds to an ignition column 
$y = E_{b,2}(1+z)/4\pi R_{\rm NS}^{2} Q_{\rm nuc}$  
ranging between $y\approx~5\times10^{8}$ g cm$^{-2}$ for 
burning hydrogen with abundance X=0.7,
and $y\approx~13\times10^{8}$ g cm$^{-2}$ for X=0 (pure helium); 
here $Q_{\rm nuc} = 1.6 + 4X$ Mev nucleon$^{-1}$ 
is the nuclear energy release for a given average hydrogen fraction  
at ignition X, and z=0.31 is the 
appropriate gravitational redshift at the surface of a 1.4 M$_{\odot}$ NS.

Interestingly, \citet{intZ07} have recently classified \IGR\ as a new 
candidate ultracompact X-ray binary, explaining its low luminosity by a small 
accretion disk. In the same category we find the burster 2S~0918-549 from which 
the observation of both short bursts and a 40 min long burst \citep{intZ05} 
indeed makes this source quite similar to \IGR.
Suggesting the companion of 2S~0918-549 is probably a helium white dwarf, 
\citet{intZ05} showed that its long burst is consistent with pure helium ignition 
and explained the different burst durations as related to the ignition
thickness and to changes in the composition of the layer between the bursts.
However, the composition of the accreted material in \IGR\ may not 
necessarily be pure helium.
The presence of some hydrogen may indeed explain 
what distinguishes burst 2 from the long burst of 2S~0918-549:
its relative weakness with respect to the Eddington limit, while 
the long burst from 2S~0918-549 did apparently reach this limit, and 
its relatively long rise time.

As a matter of fact, the inferred mass accretion rate of \IGR\ is near the value 
where the accumulating hydrogen transitions from unstable burning at low 
accretion rates, to stable burning (via the Hot CNO cycle) at higher accretion 
rates \citep[e.g.][]{s04}. 
When stable, the hydrogen burning steadily accumulates a thick helium shell 
that eventually ignites. 
Indeed, the lower limit to the burst 2 recurrence time of a few tens of days 
is in the range of those predicted by helium ignition models 
\citep[see Table 2 of][]{CB00}.
Pure helium bursts of such thick columns also lead to long burst durations 
\citep[e.g.,][]{cum06,Peng}.  
Moreover, \citet{CB00} show that for pure helium ignition, the ignition column 
is very sensitive to the accretion rate. In particular, the transition to 
unstable hydrogen burning can be quite sudden, leading to short mixed 
H/He bursts. 
As shown by \citet{CoopN} both energetic pure helium flashes and weak hydrogen 
flashes may occur near the transition. These weak hydrogen bursts (undetectable 
because their peak luminosity is lower than the persistent luminosity) 
contribute to the building 
of the deep layer of nearly pure helium and of which they may trigger the 
ignition if the mass of helium is sufficiently large.
Our burst 1 is indeed similar to a $\sim\,10$~s decay helium burst triggered by 
hydrogen ignition as simulated by \citet{Peng} sedimentation model at low 
accretion rates (see Fig. 6 of \citet{Peng}).

We conclude that 
the long burst of \IGR\ results from the ignition of a large helium pile 
beneath a steady hydrogen burning shell,
while the short event occured at a slightly lower accretion rate where 
a weak hydrogen flash prematurely triggered a mixed H/He burning burst.

\acknowledgements 
JC acknowledges financial support from ESA-PRODEX. 
MF acknowledges the French Space Agency (CNES) for financial support.

\end{document}